\documentstyle[12pt,twoside,fleqn,espcrc1,epsfig]{article}

\newcommand{\AmS}{{\protect\the\textfont2
  A\kern-.1667em\lower.5ex\hbox{M}\kern-.125emS}}

\title{What we do understand of Colour Confinement}

\author{A. Di Giacomo\address{Dipartimento di Fisica Universit\`a di Pisa
and INFN\\
        Via Buonarroti 2, 56100 Pisa, Italy}%
}

\begin{document}
\maketitle

\begin{abstract}
A review is presented of what we understand of colour confinement in QCD.
Lattice formulation provides evidence that QCD vacuum is a dual
superconductor: the chromoelectric field of a $q\bar q$ pair is constrained by
dual Meissner effect into a dual Abrikosov flux tube and the static potential
energy is proportional to the distance.
\end{abstract}

\section{Introduction}
Lattice\cite{wil} formulation is a gauge invariant regulator of non abelian
gauge
theories. Numerical simulations on the lattice produce from first principles
regulated correlators of physical quantities.

Simulations can be used to compute quantities involving low energy modes, which
are out of reach of perturbation theory, such as weak interaction matrix
elements, masses, matrix elements of operators in the light cone expansion. The
typical problems encountered in this ``phenomenological'' use of lattice are the
removal of the cut off (renormalization), and limitations in computer power.

Simulations can also be used to test theoretical ideas and to investigate the
structure of the theory. An example of investigation of ``theoretical'' type is
the study of the mechanism of confinement. The typical difficulty in this
approach is to have good theoretical ideas to
test numerically, and possibly to falsify from the first principles.

Apart from confinement itself there are a few fundamental issues at the
background of our understanding of QCD. Among them
\begin{itemize}
\item[1)]
The $1/N_c\to 0 $ limit. The conjecture\cite{thooft} is that the
basic properties of a gauge theory, e.g. confinement, are already
contained in the limit in which the number of colours $N_c$ goes
large, with $g^2 N_c$ fixed. Corrections ${\cal O}(1/N_c)$ can be
treated as a small perturbation. A consequence of this conjecture
is that also quark loops can be viewed as a small perturbation,
apart from their effect on the scale of the theory. Indeed apart
from the loop with two vertices which is proportional to $g^2
N_f\sim N_f/N_c\sim 1$, and which enters in the $\beta$ function,
loops with more vertices have additional factors $1/N_c$ and are
negligible. According to this conjecture also the mechanism of
confinement and the corresponding order parameter should then be
marginally affected by the presence of quarks.
\item[2)] Understanding the ground state is also important to understand why
perturbation theory works at small distances.
Perturbative quantization describes interaction of quarks and gluons, and the
ground state is the Fock vacuum. Quarks and gluons are not observed in nature,
and the Fock vacuum is certainly not the ground state. This reflects in the
renormalized perturbation expansion as a lack of convergence, even in the sense
of asymptotic expansion\cite{mul}.
\end{itemize}
\section{Confinement in Nature.}
Colour is confined in nature. The expected ratio
of abundance of quarks $n_q$ to abundance
of nucleons $n_p$ is in the standard cosmological model\cite{Okun}
\begin{equation}
\frac{n_q}{n_p} \simeq 10^{-12} \ . \label{eq:e1}\end{equation}
The experimental upper limit is
\begin{equation}
\frac{n_q}{n_p}\leq 10^{-27} \ , \label{eq:e2}\end{equation}
coming from Millikan like experiments on $\sim 1\,g$ of matter.

The estimate (\ref{eq:e1}) is conservative. If we assume no confinement and $T$
is the temperature at which quarks decouple their effective mass $m_q$ is
$\sim T$. The reactions
\[ q + \bar q \to {\rm mesons}\qquad
q + q \to \bar q +\,{\rm baryons}\]
are esothermic: let $\sigma$ the corresponding cross section and
$\sigma_0 \equiv \lim_{v\to 0} \sigma v$. Then quarks will decouple when
\begin{equation}
n_q \sigma_0 = G_N^{1/2} T^2 \ . \label{eq:e3}\end{equation}
Since $n_\gamma \sim T^3$ this implies
\begin{equation}
\frac{n_q}{n_\gamma} \sim \frac{ G_N^{1/2}}{T \sigma_0} =
\frac{10^{-18}}{m_p m_q \sigma_0} \ , \label{eq:e4}\end{equation}
$\sigma_0\sim m_\pi^{-2}$. The ratio (\ref{eq:e1}) corresponds to $m_q = T =
10\,GeV$.

The factor $10^{-15}$ between the observation and the expectation cannot be
explained by fine tuning of a small parameter. Like the experimental limit on
the resistivity of a superconductor, it can only be explained in terms of
symmetry.

A suggestive idea in that direction is that vacuum is a dual superconductor
\cite{Mandelstam}.
The chromoelectric field between a $q$ $\bar q$ pair is constrained by dual
Meissner effect into an Abrikosov flux tube with energy proportional to the
distance.

A relativistic version of the free energy of a superconductor, which is the
analog of effective action in field theory, is
\begin{equation}
G = -\frac{1}{4} F_{\mu\nu} F_{\mu\nu} +
(D_\mu\varphi)^* (D_\mu\varphi) + V(\varphi) \ , \label{eq:e5}\end{equation}
where
\begin{equation}
D_\mu\varphi = (\partial_\mu - i q A_\mu)\varphi\label{eq:e6}\end{equation}
is the covariant derivative and
\begin{equation}
V(\varphi) = \mu^2 \varphi^*\varphi - \frac{\lambda}{2}(\varphi^*\varphi)^2
\label{eq:e7}\end{equation}
is the effective potential.
$\mu^2$ and $\lambda$ are funtions of the temperature, and $\mu^2(T) > 0$ in
the superconducting phase, where the potential has a mexican hat shape.

Putting $\varphi = f e^{i q \theta}$, with $f > 0$ gives
$
D_\mu \varphi = - i q (A_\mu - \partial_\mu\theta) f e^{i q \theta}
$.
Under a gauge transformation $A_\mu \to A_\mu + \partial_\mu \Lambda$
\begin{equation}
\partial_\mu \theta \to \partial_\mu \theta + \partial_\mu\Lambda
\label{eq:e9}\end{equation}
and $\tilde A_\mu = A_\mu - \partial_\mu\theta$ is gauge invariant.
Moreover
\begin{equation}
\tilde F_{\mu\nu} = \partial_\mu\tilde A_\nu - \partial_\nu\tilde A_\mu=
F_{\mu\nu}\label{eq:e10}\end{equation}
and the free energy can be rewritten as
\begin{equation}
G = -\frac{1}{4}\tilde F_{\mu\nu}\tilde F_{\mu\nu} +
q^2 f^2 \tilde A_\mu\tilde A_\mu +
\partial_\mu f \partial_\mu f + \mu^2 f^2 - \frac{\lambda}{2} f^4
\label{eq:e11} \ . \end{equation}
The equations of motion are
\begin{equation}
\partial_\mu \tilde F_{\mu\nu} + q^2 f^2 \tilde A_\nu = 0 \ , \qquad
(\partial^2 + \mu^2) f = 2 \lambda f^3
\label{eq:e12}\end{equation}
A static solution in the gauge $A_0=0$ has $\vec E = 0$, $f = \tilde f \equiv
\sqrt{\mu^2/2\lambda}$ and eq.(\ref{eq:e12}) reads
\begin{eqnarray}
\vec\nabla\wedge\vec H + q^2 {\tilde f}^2 \vec{\tilde A} &=& 0
\label{eq:e13a} \ , \\
\nabla^2\vec H - q^2 {\tilde f}^2 \vec H &=& 0  \ . \label{eq:e13b}
\end{eqnarray}
Eq.~(\ref{eq:e13a}) means that also in the absence of electric field there is a
permanent current, or that $\sigma = \infty$. Eq.~(\ref{eq:e13b}) is Meissner
effect: the field $\vec{\tilde A}$ penetrates by a length $(q \tilde f)^{-1}$.
At large distance from the center of a flux tube $\vec{\tilde A} = 0$ or
\begin{equation}
e \oint \vec{\tilde A} d\vec x = e \Phi(B) = 2\pi n \ ,
\label{eq:e14}\end{equation}
which is the Dirac quantization condition.
Abrikosov flux tubes have monopoles at their ends.

Those phenomena are a consequence of symmetry\cite{wei}:
the order parameter is $f = \langle | \varphi| \rangle$, or the non vanishing
v.e.v. of a charged
operator. For dual superconductivity the signal of the phase should be the
v.e.v. $\langle\mu\rangle$ of an operator carrying magnetic charge.
\section{Phenomenology of confinement on the lattice.}
Lattice produces evidence for confinement. Wilson loops, defined as parallel
transport along square contours in space time, provide the static force between
$q\bar q$ pairs, in the limit of large $T$
\begin{equation}
W(R,T) \mathop{\simeq}_{T \to \infty}
\exp(-\sigma V(R) T)\label{eq:e15}\end{equation}
The area law observed in lattice gauge theory\cite{creu}
\begin{equation}
W(R,T) \simeq \exp(-\sigma R T)\label{eq:e16}\end{equation}
means
\begin{equation}
V(R) = \sigma R \label{eq:e17}\end{equation}
or that an infinite amount of energy is needed to pull the two particles at
infinite distance from each other. $\sigma$ is the string tension, related
to the slope of Regge trajectories.

Also chromoelectric flux tubes between $q\,\bar q$ pairs are observed, with
transverse size $\sim 0.5\,{\rm fm}$\cite{hay,adg1}. The colour orientation of
the chromoelectric field inside them can also be studied\cite{adg1,gree}.

Finally the collective modes of the string formed by the flux tube can be
analysed\cite{Gliozzi}.

All these facts support the picture of confinement as due to dual
superconductivity of vacuum.
A microscopic understanding is however needed. In particular monopoles
which condense in the vacuum have to be identified.

\section{Monopoles}
In QED, which is a $U(1)$ gauge theory, magnetic charges are omitted,
since they are not observed in nature. As a consequence the general solution
of Maxwell's equations can be given in terms of a vector potential $A_\mu$.
The field strength tensor
\begin{equation}
F_{\mu\nu} = \partial_\mu A_\nu - \partial_\nu A_\mu\label{eq:e18}
\end{equation}
obeys the equations
\begin{equation}
\partial_{\mu} F_{\mu \nu} = j_{\nu} \ .
\end{equation}
The absence of magnetic charges indentically follows from eq.(\ref{eq:e18}).
The dual tensor $F^*_{\mu\nu} \equiv \frac{1}{2}
\varepsilon_{\mu\nu\rho\sigma}F^{\rho\sigma}$ is, by virtue of eq.
(\ref{eq:e18}) identically conserved:
\begin{equation}
\partial_\mu F^*_{\mu\nu} = 0 \ . \label{eq:e19}\end{equation}
Eq.~(\ref{eq:e19}) is known as Bianchi identity.

The only way to have a monopole and to preserve Bianchi identity is
\cite{Dirac} to introduce a singularity, and consider the monopole as the
end point of an infinitely thin solenoid (Dirac string), which can be
made invisible if the parallel transport of any charge $q$ around it
is trivial or if
\begin{eqnarray}
\exp \left(iq \oint \vec{A} \cdot \,d \vec{x}\right) = 1 \ .
\label{eq:e20}
\end{eqnarray} 
The line integral is intended on a path which encircles the string and is
equal to the magnetic flux, or to $4 \pi$ the magnetic charge of the monopole.
Eq.~(\ref{eq:e20}) implies $4 \pi q M = 2 n \pi$ or $q M = \frac{n}{2}$, which
is the celebrated Dirac quantization condition. As a consequence the theory
becomes compact.

In non abelian gauge theories, in the familiar multipole expansion the
monopole term obeys abelian equation of motion, has a Dirac string and a
number of independent abelian magnetic charges which is $N-1$ for the gauge
group $SU(N)$\cite{Coleman}.
The 't Hooft-Polyakov\cite{'tHooft,Polyakov} monopole
of the $SO(3)$ Georgi-Glashow model obeys this classification. 
\section{Monopoles in QCD}
To understand the monopoles in QCD we shall phrase the classification
of ref.~\cite{Coleman} in the language of ref.~\cite{'tHooft2}, or in terms
of ``abelian projections''. We shall refer to $SU(2)$
for simplicity: extension to $SU(N)$ is trivial\cite{adg2}.

Let $\vec{\phi}(x) \cdot \vec{\sigma}$ be any local operator belonging to the
adjoint representation. We define $\hat{\phi}(x) = \frac{\vec{\phi}(x)}
{\left|\vec{\phi}(x)\right|}$. $\hat{\phi}$ is well defined except at zeros
of $\vec{\phi}(x)$. Consider the field strength tensor\cite{'tHooft}
\begin{eqnarray}
F_{\mu \nu}(x) = \hat{\phi} \cdot \vec{G}_{\mu \nu}(x) - \frac{1}{g}
\hat{\phi}(x) \cdot \left( D_{\mu} \hat{\phi}(x)\wedge D_{\nu}
\hat{\phi}(x) \right)
\label{eq:e21}
\end{eqnarray}
where $\vec{G}_{\mu \nu}(x) = \partial _{\mu} \vec{A} _{\nu} -
\partial _{\nu} \vec{A} _{\mu}$ and $D _{\mu} = \partial _{\mu} - g  \vec{A}
\wedge$ is the covariant derivative. The coefficient of the second
term in eq.~(\ref{eq:e21}) is chosen in such a way that the quadratic term
$\vec{A}_{\mu} \wedge \vec{A}_{\nu}$ cancels with the first term.
Both terms are gauge invariant under regular gauge tranformations.

A gauge transformation $U(x)$ which brings $\hat{\phi}$ along the 3 axis
$\hat{\phi} = (0,0,1)$, and diagonalizes $\vec{\phi}(x) \cdot \vec{\sigma}$ 
is called an abelian projection. After abelian projection
\begin{eqnarray}
F_{\mu \nu} = \partial _{\mu} A^3_{\nu} -  \partial _{\nu} A^3_{\mu}   
\end{eqnarray}
is an abelian field. This holds in all points where $U(x)$ is regular.
Defining the dual tensor $F_{\mu \nu}^{\star}$ as
\begin{eqnarray}
F_{\mu \nu}^{\star} = \frac{1}{2} \epsilon _{\mu \nu \rho \sigma}
F_{\rho \sigma} \ ,
\end{eqnarray}
the magnetic current $j_{\mu}^M$ is defined as
\begin{eqnarray}
\partial_{\mu} F_{\mu \nu}^{\star} = j_{\nu}^M
\end{eqnarray}
and is identically conserved. It is zero except at the singular points of
$U(x)$, where monopoles can appear. Thus
\begin{eqnarray}
\partial_{\mu} j^{M} _{\nu} = 0
\end{eqnarray}
defines a magnetic $U(1)$ symmetry of the theory. It is not a subgroup of
the gauge group since both $F_{\mu \nu}$ and $j_{\mu}^M$ are colour singlet.
If the vacuum is not invariant under that $U(1)$ symmetry, monopoles condense
like the Cooper pairs and there is dual superconductivity.

Notice that
\begin{enumerate}
\item there is a magnetic $U(1)$ symmetry for each operator $\vec{\phi}(x)$
in the adjoint representation.
\item Under the abelian projection $U(x)$, due to singularities, the field
strength tensor $\vec{G}_{\mu \nu}$ acquires a singular component
\begin{eqnarray}
\vec{G}_{\mu \nu} \to U \vec{G}_{\mu \nu} U^{-1} +
\vec{G}_{\mu \nu} ^{sing} \ .
\end{eqnarray}
The regular part can have monopole sources. $\vec{G}_{\mu \nu} ^{sing}$
describes Dirac strings starting from the monopoles. 
\end{enumerate}

The strategy will then be to detect condensation of different monopole
species by measuring with numerical simulations across the deconfining
transition a ``disorder'' parameter, which detects dual superconductivity.
The usual order parameter is the Polyakov line. Our disorder parameter
will be zero in the deconfined phase, where the order parameter is
non zero, and different from zero in the confined phase, where it is zero.
The concept of disorder parameter is typical of systems which admit a dual
description\cite{Kadanoff}. They usually have extended structures
with non trivial topology (monopoles in QCD), which condense in the disordered
phase. In a dual description these structures are described by local fields,
and the definition of order and disorder is interchanged. Before giving the
results and a few details on how they have been obtained, we shall present
the expectations.
\section{Expectations vs. results}
As we have seen dual superconductivity of the vacuum is not a well defined
concept. There are infinitely many choices for the operator $\vec{\phi}(x)$,
and for each of them ther can or can not be condensation. What is the good
choice, if any?
\begin{enumerate}
\renewcommand{\theenumi}{\Alph{enumi}}
\item There is club of practitioners of the ``maximal abelian projection'',
saying that their choice is better than others. In fact with this choice the
abelian field ``dominates'' the configurations, in particular the part of
it which is produced by monopoles. This can prove convenient to attempt a
construction of effective lagrangeans, but in principle does not preclude
any pattern of symmetry.
\item There is a conjecture that all abelian projections are
equivalent\cite{'tHooft}. 
\end{enumerate}
Most probably both attitudes reflect our imperfect knowledge of the symmetry
of the disordered phase.

Discriminating between (A) and (B) is possible on the lattice. The results
that we have obtained by a systematic study of dual superconductivity
show unambiguously 
\begin{enumerate}
\item That confinement is a transition from normal to dual superconductor
ground state. This is strong evidence that the mechanism of confinement is
indeed dual superconductivity of the vacuum.
\item A few different abelian projections have been analyzed. All of them
show the same behaviour. The scenario (B) seems to be true.
\end{enumerate}

This is reassuring for the validity of the mechanism itself.
If only one abelian projection would show dual superconductivity
only the particles with non zero charge with respect to that
$U(1)$ could be confined:
there exist states for which that charge is zero, e.g. the gluon which is
parallel to $\hat{\phi}$. Moreover the colour direction of the
electric field in the flux tubes observed in the lattice should also be
parallel to $\hat{\phi}$, being the electric partner of the magnetic $U(1)$
of the monopoles. This has been shown to be not true\cite{adg1,gree}.
Both these facts are naturally explained if the scenario (B) is at work.

We conclude by giving a few details on the technique used to detect dual
superconductivity. The technique has been checked in many well known
systems showing order disorder duality versus traditional
descriptions\cite{tutti}.
\section{The disorder parameter}
An operator $\mu$ is constructed which carries non zero magnetic charge
with respect to the $U(1)$ under study.
Finite temperature is realized
on the lattice by the usual thermodynamical recipe of having euclidean
time running from $0$ to $1/T$ (the temperature), with periodic boundary
conditions for barions, antiperiodic for fermions. On a lattice this is
done by using a size $N_s^3 \times N_t$, with $N_s \gg N_t$ and $a(\beta) N_t
= 1/T$ ($\beta = 2N/g^2$). By renormalization group arguments
$a(\beta) \simeq \frac{1}{\Lambda _L} \exp(- b_0 \beta)$,
or $T = \frac{\Lambda _L}{N_t} \exp( b_0 \beta)$.

The technique used to construct $\mu$ is inspired to ref.~\cite{Kadanoff}
and~\cite{Swieca} and is a complicated version of the simple formula for
translations in elementary quantum mechanics
\begin{eqnarray}
e^{ipa} | q \rangle = | q + a \rangle \ .
\end{eqnarray}
In the Schr\"odinger representation, the field $\vec{A}(x,t)$ plays the r\^ole
of $x$, the conjugate momentum $\vec{\Pi}(\vec{x},t)$ the r\^ole of $p$ and
\begin{eqnarray}
\mu |\vec{A}(\vec{x},t) \rangle \equiv \exp^{\left( i \int \,d^3 x \vec{\Pi}
(\vec{x},t) \cdot \vec{\bar{A}}( \vec{x}- \vec{y}) \right)}
| \vec{A}(\vec{x},t) \rangle = | \vec{A}(\vec{x},t) + \vec{\bar{A}}(\vec{x} -
\vec{y}) \rangle \ .
\label{eq:e31}
\end{eqnarray} 
If $\vec{\bar{A}}$ is the field of a monopole at $\vec{y}$, $\mu$ is indeed
the creation operator for a monopole, at site $\vec{y}$ and at time $t$. When
inserted in the Feynman integral the operator is nothing but a linear term
in the conjugated momentum added to the lagrangean and hence a shift of $
\vec{\Pi}$ at
time $t$. Care is needed to adapt the definition (\ref{eq:e31}) to a compact
system, in a form which does not depend on the choice of the gauge for the
classical field $\vec{\bar{A}}$. This can be done, and the result is, as
sketched above\cite{adg2,tutti}
\begin{eqnarray}
\langle \mu \rangle = \frac{Z[S + \Delta S]}{Z[S]} \ ,
\end{eqnarray}
with $\Delta S$ different from zero on a hyperplane at constant $x_0$.

\begin{figure}
\begin{minipage}[htb]{7.8cm}
\centering{\epsfig{figure=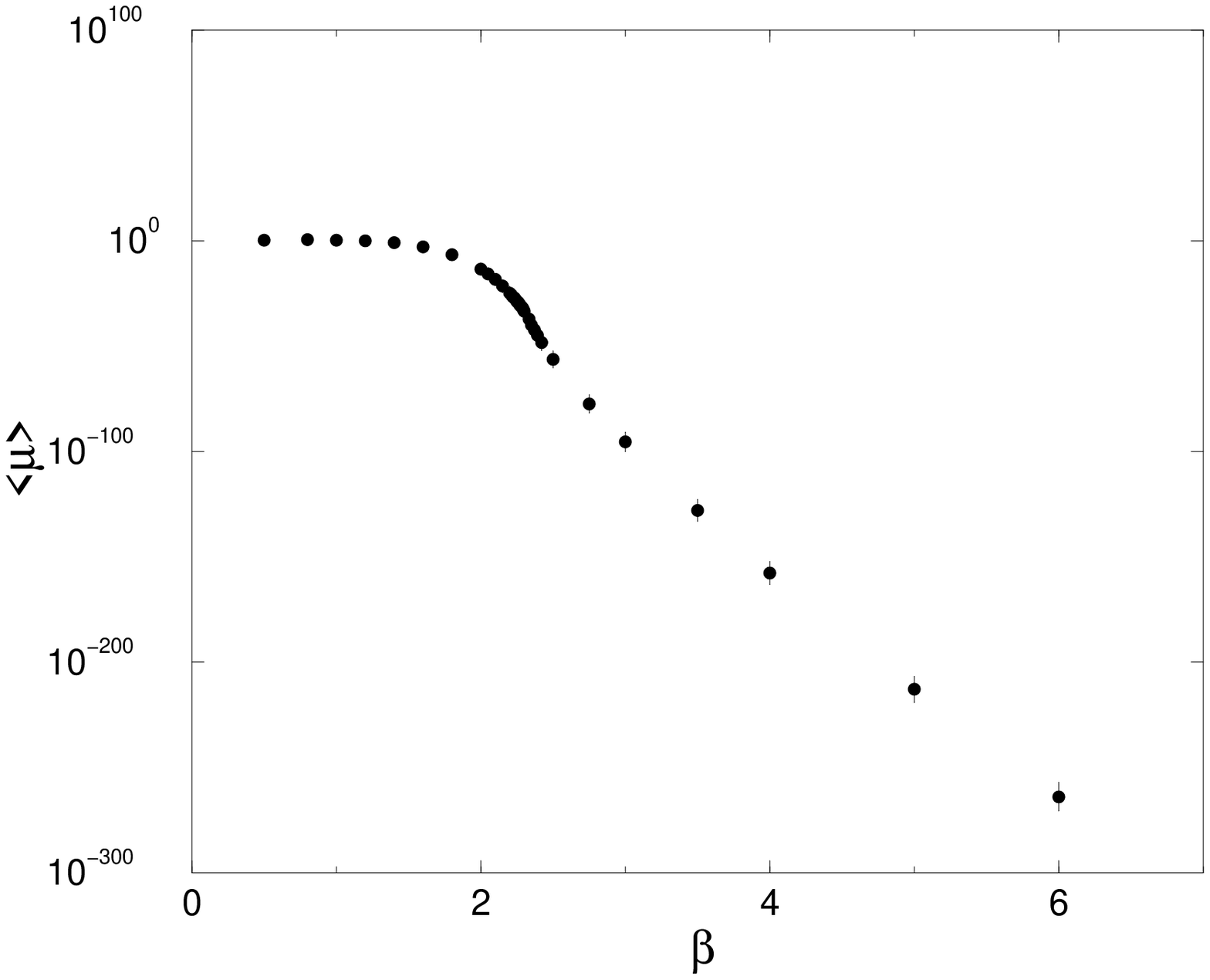, width=7.8cm}}  
\caption{$\rho$ vs. $\beta$ for $SU(2)$ gauge theory. Plaquette projection,
lattice $16^3 \times 4$.}
\label{fig1}
\end{minipage}
\hfill
\begin{minipage}[htb]{7.8cm}
\centering{\epsfig{figure=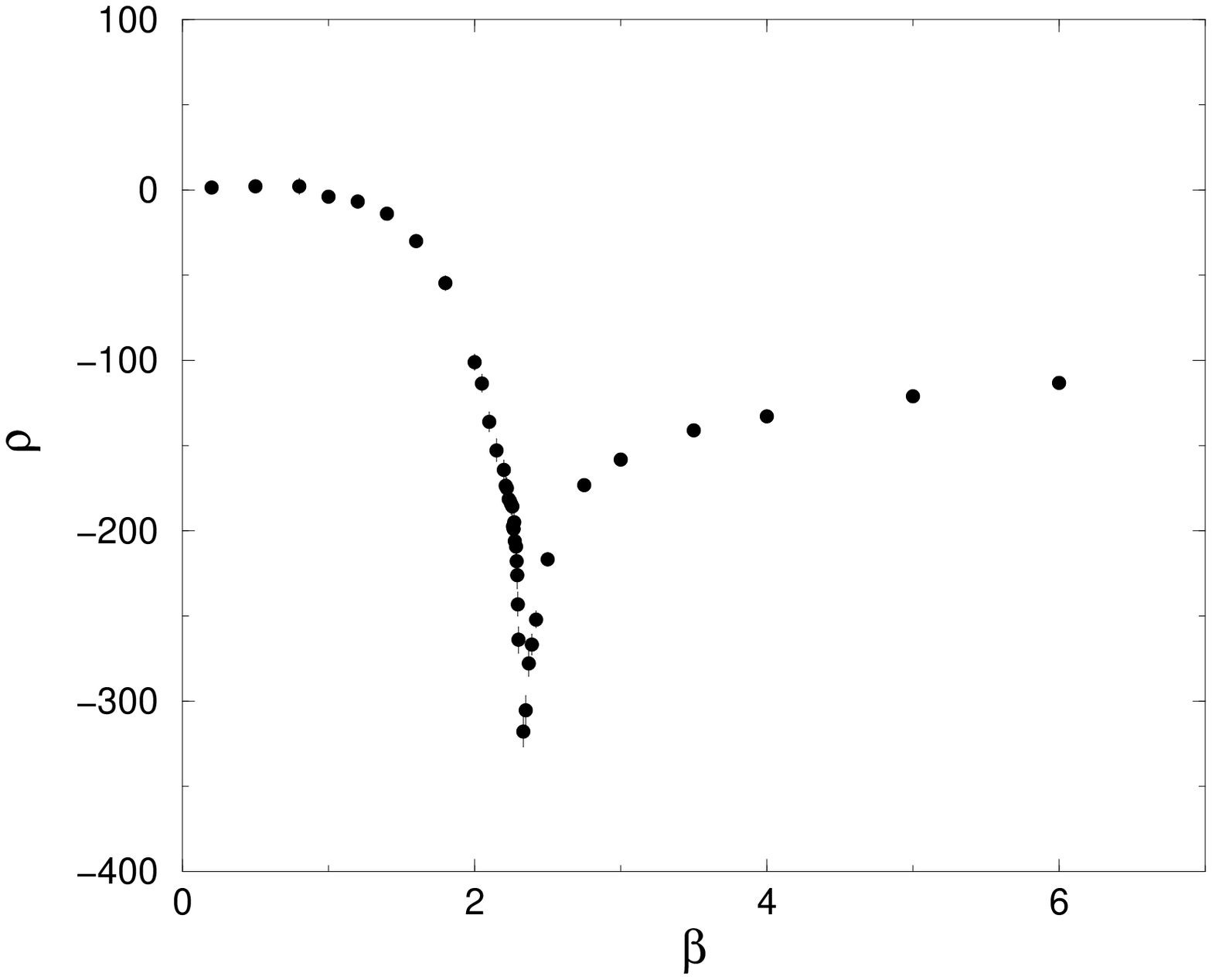, width=7.8cm}} 
\caption{$\langle \mu \rangle$ vs. $\beta$ for $SU(2)$ gauge theory.
Plaquette projection, lattice $16^3 \times 4$.}
\label{fig2}
\end{minipage}
\end{figure}

\begin{figure}
\begin{minipage}[htb]{7.6cm}
\centering{\epsfig{figure=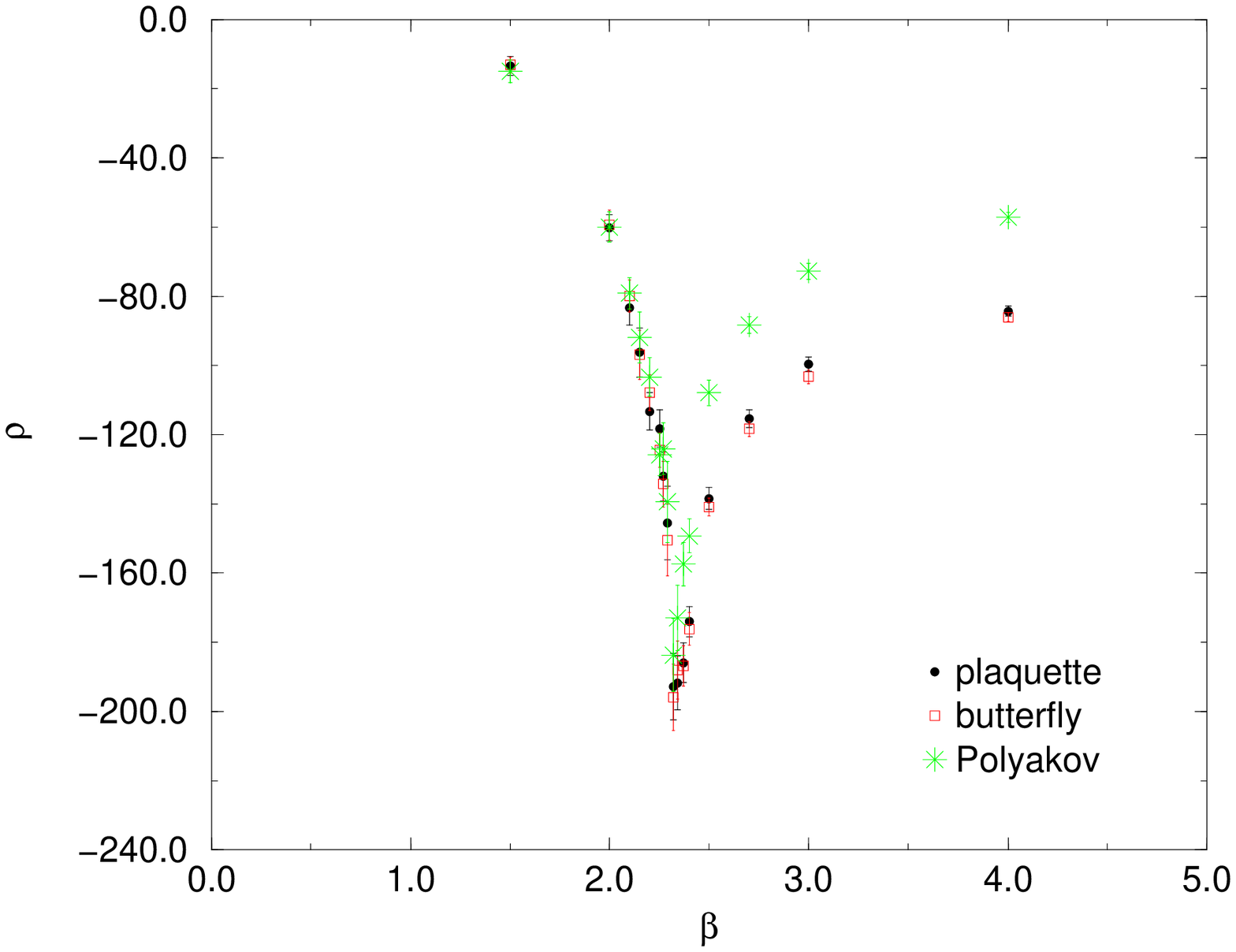, width=7.6cm}}  
\caption{$\rho$ vs. $\beta$ for different abelian projections. $SU(2)$ gauge
theory, lattice $12^3 \times 4$.}
\label{fig3}
\end{minipage}
\hspace{0.3cm}
\begin{minipage}[hb]{7.5cm}
~\vspace{1mm}
\centering{\epsfig{figure=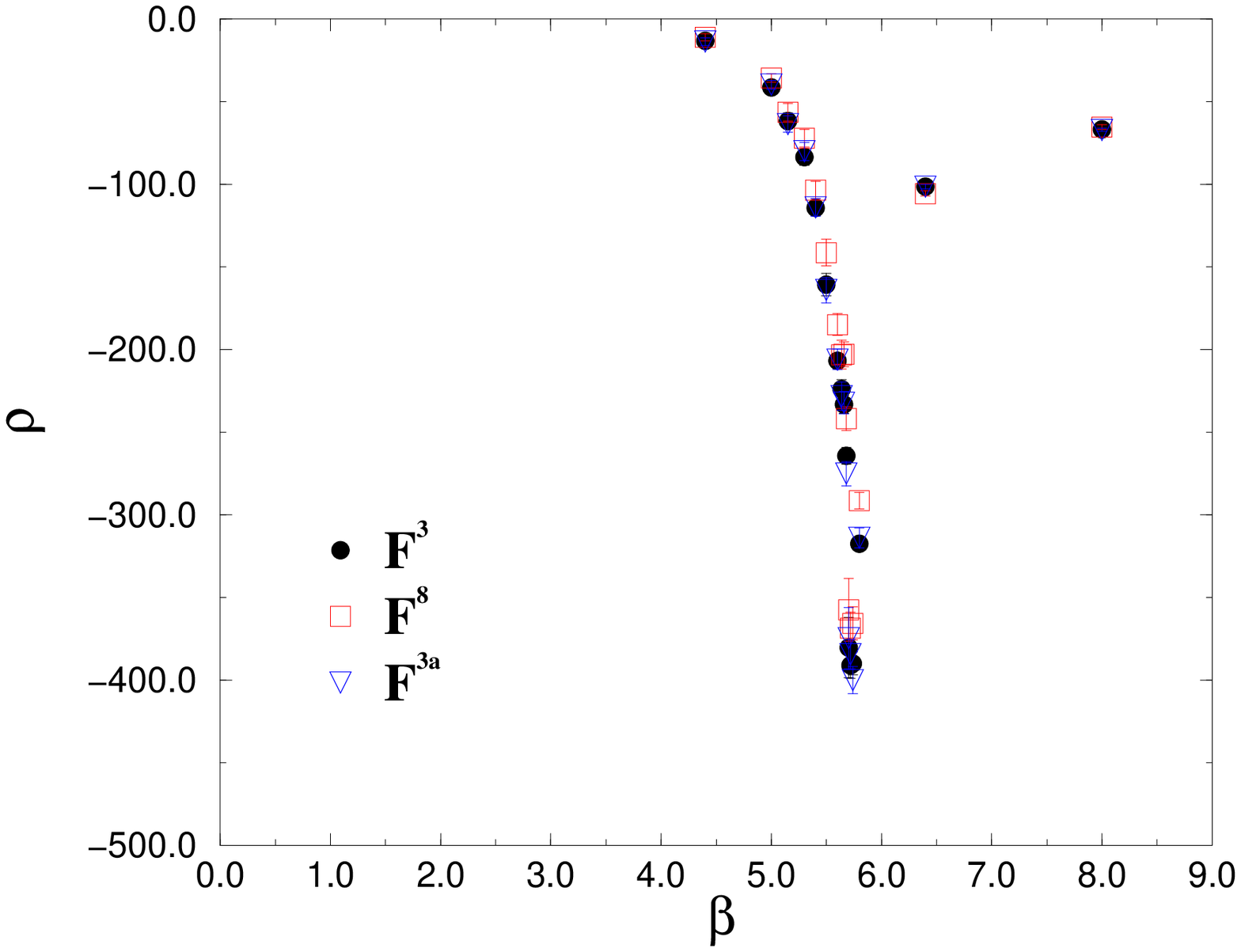, width=7.5cm}} 
\caption{$\rho$ vs. $\beta$ for different abelian generators of $SU(3)$.
Polyakov projection, lattice $12^3 \times 4$.}
\label{fig4}
\end{minipage}
\end{figure}

Being the exponential of a sum on $N_s^3$ sites $\langle \mu \rangle$ is
subject to strong fluctuations. Numerically it is better to measure
the quantity
\begin{eqnarray}
\rho = \frac{\mbox{d}}{\mbox{d}\beta} \log \langle \mu \rangle = 
\langle S \rangle _S -  \langle S + \Delta S \rangle _{S + \Delta S}
\end{eqnarray}
and to reconstruct $\langle \mu \rangle$ as
\begin{eqnarray}
\langle \mu \rangle = \exp \left[ \int_0^{\beta}
\rho(\beta^{\prime}) \,d \beta^{\prime} \right] \ .
\end{eqnarray}
A typical shape of $\rho$ is shown in fig.~\ref{fig1}, one of $\langle
\mu \rangle$ in fig.~\ref{fig2}. As is well known\cite{Lee} $\langle \mu
\rangle$ as an analytic function of $\beta$, can not vanish indentically in
the deconfined phase if the number of degrees of freedom is finite.
An extrapolation to $N_s \to \infty$ must be done by finite size analysis.
Fig.~\ref{fig3} shows that a few different abelian projections behave
in the same way. Fig.~\ref{fig4} shows that two different monopole
species of $SU(3)$ also behave in the same way.

\noindent

\begin{figure}
\begin{minipage}[htb]{7.8cm}
\centering{\epsfig{figure=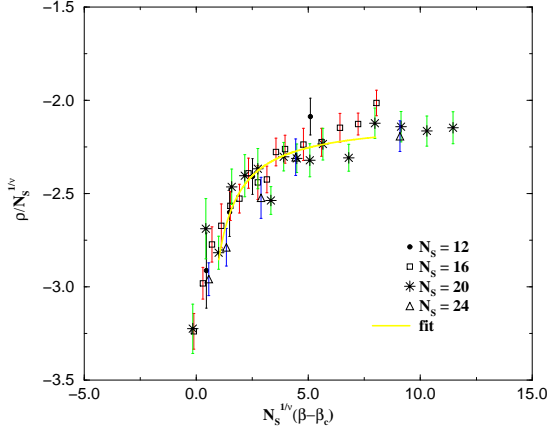, width=8cm}}  
\caption{Quality of the scaling for $SU(2)$ gauge theory.
Plaquette projection, $N_t = 4$.}
\label{fig5}
\end{minipage}
\hfill
\begin{minipage}[htb]{7.8cm}
\centering{\epsfig{figure=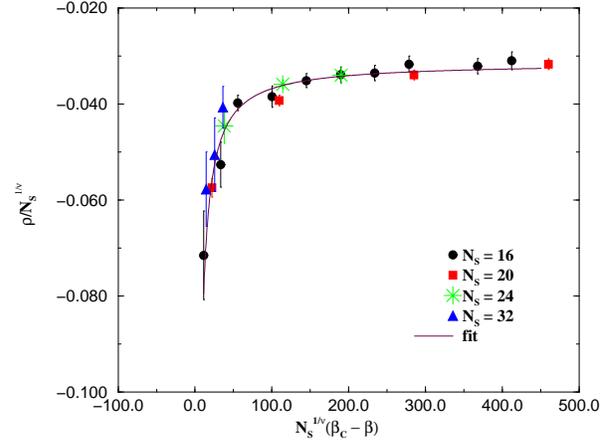, width=7.8cm}} 
\vspace{0.2cm}
\caption{Quality of the scaling for $SU(3)$ gauge theory.
Polyakov projection, $N_t = 4$.}
\label{fig6}
\end{minipage}
\end{figure}

As for the extrapolation to $N_s \to \infty$ three different ranges
of $\beta$ are considered:
\begin{enumerate}
\item $\beta \gg \beta_c$: there perturbative theory is at work and $\rho
\simeq - c N_s + d$, with $c$ a positive constant. As $N_s \to \infty$
$\rho  \to - \infty$ and $\langle \mu \rangle = 0$.
\item $\beta \ll \beta_c$: there $\rho$ has a finite limit as $N_s \to \infty$,
and $\langle \mu \rangle$ is finite.
\item $\beta \simeq \beta_c$: there we expect $\langle \mu \rangle \simeq
\left( \beta_c - \beta \right) ^{\delta}$. From dimensional analysis
\begin{eqnarray}
\langle \mu \rangle = N_s^{\delta} \Phi \left( \frac{a}{\xi}, \ 
\frac{\xi}{N_s}, \ \frac{N_t}{N_s} \right) \ .
\end{eqnarray}
Where $\xi \gg a$, and if $N_s \gg N_t$
\begin{eqnarray}
\langle \mu \rangle = N_s^{\delta} \Phi \left( 0, \ 
\frac{\xi}{N_s}, \ 0 \right) \ ,
\end{eqnarray}
and since $\xi \simeq \left( \beta_c - \beta \right)^{- \nu}$, we get
the scaling law
\begin{eqnarray}
\frac{\rho}{N_s^{1/\nu}} = f \left( N_s^{1/\nu}(\beta_c - \beta)\right) \ .
\end{eqnarray}
The quality of the scaling is shown in fig.~\ref{fig5} for $SU(2)$ and
in fig.~\ref{fig6} for $SU(3)$. It gives a determination of
the critical indices $\nu$, $\delta$, and of $\beta_c$.
For $SU(2)$ we get
\begin{eqnarray}
\nonumber
\hspace{4cm}
\begin{array}{l}
\nu = 0.63(5)\\
\beta_c = 2.30(2) \qquad N_t = 4 \\
\delta = 0.20(8)
\end{array}
\ .
\end{eqnarray}
$\nu$ and $\beta _c$ are in agreement within errors  with independent
determinations, and $\nu$ indicates a second order phase transition.

For $SU(3)$
\begin{eqnarray}
\nonumber
\hspace{4cm}
\begin{array}{l}
\nu = 0.33(2)\\
\beta_c = 5.70(3) \qquad N_t = 4 \\
\delta = 0.54(4)
\end{array}
\ ,
\end{eqnarray}
indicating that the transition is first order.
\end{enumerate}

\noindent
\noindent

The method used has been tested on a number of known systems and understood
in its details. The result show beyond any reasonnable doubt that dual
superconductivity occurs, in different abelian projections, in connection
with confinement.

\vspace{0.5cm}
The part of the work reported due to our group has been done in collaboration
with L. Del Debbio, G. Paffuti, P. Pieri, B. Lucini, D. Martelli in the last
few years. Their contribution was determinant to the results.

\end{document}